\documentclass[screen,sigconf]{acmart}

\AtBeginDocument{%
  \providecommand\BibTeX{{%
    \normalfont B\kern-0.5em{\scshape i\kern-0.25em b}\kern-0.8em\TeX}}}

\copyrightyear{2025}
\acmYear{2025}
\setcopyright{rightsretained}
\acmConference[ASSETS '25]{The 27th International ACM SIGACCESS Conference on Computers and Accessibility}{October 26--29, 2025}{Denver, CO, USA}
\acmBooktitle{The 27th International ACM SIGACCESS Conference on Computers and Accessibility (ASSETS '25), October 26--29, 2025, Denver, CO, USA}
\acmDOI{10.1145/3663547.3759766}
\acmISBN{979-8-4007-0676-9/2025/10}

\usepackage{todonotes}
\usepackage{soul}  
\usepackage{enumitem}

\setcopyright{none}

\begin{document}

\title{\textit{Phoenix}: A Novel Context-Aware Voice-Powered Math Equation Workspace and Editor}

\author{Kenneth Ge}
\authornote{All three authors contributed equally to this research.}
\orcid{0009-0000-5044-4433}
\affiliation{%
  \institution{Assistivity}
  \city{Edgemont}
  \state{NY}
  \country{USA}
  \postcode{10583}
}
\email{kennethkouge@gmail.com}

\author{Ryan Paul}
\authornotemark[1]
\orcid{0009-0004-5440-1827}
\affiliation{%
  \institution{Assistivity}
  \city{Brampton}
  \state{ON}
  \country{Canada}
  \postcode{L7A 4M4}
}
\email{paulryan229@gmail.com}

\author{Priscilla Zhang}
\authornotemark[1]
\orcid{0009-0005-4112-766X}
\affiliation{%
  \institution{Assistivity}
  \city{Villanova}
  \state{PA}
  \country{USA}
  \postcode{19073}
}
\email{zhang.priscilla34@gmail.com}

\author{JooYoung Seo}
\orcid{0000-0002-4064-6012}
\affiliation{%
 \department{School of Information Sciences}
  \institution{University of Illinois Urbana-Champaign}
  \city{Champaign}
  \state{Illinois}
  \country{USA}
  \postcode{61820}
  }
\email{jseo1005@illinois.edu}

\renewcommand{\shortauthors}{Ge et al.}

\begin{abstract}

Writing mathematical notation requires substantial effort, diverting cognitive resources from conceptual understanding to documentation mechanics, significantly impacting individuals with fine motor disabilities (FMDs). Current limits of speech-based math technologies rely on precise dictation of math symbols and unintuitive command-based interfaces. We present a novel voice-powered math workspace, applying neuroscience insights to create an intuitive problem-solving environment. To minimize cognitive load, we leverage large language models with our novel context engine to support natural language interaction. Ultimately, we enable fluid mathematical engagement for individuals with FMDs-- freed from mechanical constraints

\end{abstract}



\received{20 February 2007}
\received[revised]{12 March 2009}
\received[accepted]{5 June 2009}

\maketitle

\section{Introduction}
\label{sec:introduction}

\begin{quotation}
An injury six years ago rendered both my wrists intermittently unusable. Despite previously excelling in math and science, I experienced the limits of mathematical voice-to text software firsthand. Existing tools were unpredictable, translating my speech into linear fragments of English and mathematical notation, and failing to capture nested and multidimensional equations. Even solving simple algebra problems required extra time and energy, without features that understood commands like "divide both sides by 2." Tasks previously requiring only insight and reasoning were frequently interrupted and constrained by translational inaccuracies and mechanical obstacles. My progress slowed, not due to differences in intellect or effort, but a lack of technical tools. 
\end{quotation}

P's experiences--as a co-author on our mixed-ability design team--reflect those of many individuals with fine motor disabilities (FMDs), and directly inspired \textit{Phoenix's} development. Although foundational in STEM \cite{kristensen2024role,forde2023investigating,nctm2003mathematics,BENJACOB2019IMP}, mathematical concepts and notation pose unique accessibility challenges due to the need to manipulate specialized symbols, spatially arrange formulas, and structure multilevel equations \cite{Geary1993-xd, ge_stereomath_2024}. Such processes often require handwriting or typing, placing high demands on physical dexterity \cite{widada_overcoming_2020,wason_natural_1971,koedinger_real_2004}.  

Barriers emerge as early as preschool, where fine motor skills are strong predictors of mathematical success \cite{Flores2023-cz,Pitchford2016-bc}. Throughout schooling, understaffed disability services, rigid service protocols, and lack of faculty awareness compound previous difficulties \cite{dowrick2005postsecondary, witham_disabilities2024, noauthor_understanding_nodate}. Although scribes help accessibility, mismatches in pacing and interpretation divert attention from core concepts \cite{ressa2021scribe, paulo_2017}. Fixed mindsets around math and widespread "math anxiety" particularly harm individuals with FMDs, who may internalize a sense of inadequacy and disengage due to reinforcing cycles \cite{dong_cultural_2022, mokobane_fine_2019, khasawneh_what_2021, rozgonjuk_mathematics_2020}. 

Existing solutions require users to dictate rigid grammar structures, introducing steep learning curves and challenging navigational demands. However, neuroscience research suggests that mathematical ability comes from symbolic and spatial reasoning, and is distinct from motor ability \cite{Lee01012011, van_garderen_visualspatial_2003,vale_importance_2017, knauff_spatial_2002}. 

Building on these insights, we present \textit{Phoenix}, a voice-based math workspace designed from the ground up to minimize cognitive overhead. Similar to the mythical creature that is reborn from its ashes, \textit{Phoenix} enables users to overcome physical obstacles and find renewal in mapping mathematical ideas in a novel way. We integrate structure, context, and naturally spoken mathematics by leveraging recent advances in large language models (LLMs). We aim to transform mathematical problem solving into an intuitive and high-level process for users at all levels, from students to professionals.  We focus on the following research questions:
\begin{itemize}[topsep=0pt]
\item How can we help users with fine motor disabilities intuitively map out mathematical ideas?
\item How can we enable new interaction paradigms and enhanced understanding through robust voice input?
\item How can an editor optimized for cognitive load enable new levels of abstract problem solving?
\end{itemize}

\section{Related Work}
\label{sec:related_work}

As most FMD users prefer voice input modalities \cite{li_2022}, speech-based math editors demonstrate accessibility potential. Early systems like \textit{MathTalk} used structured pauses to convey meaning \cite{robert_1994}, while \textit{CamMath} and \textit{TalkMaths} progressed towards natural classroom-style speech \cite{cameron2007computer, wigmore_2010}. \textit{Mathifier} further introduced a grammar, albeit highly structured and code-like \cite{batlouni_2011}. Although these approaches establish voice as effective input, they introduce cognitive overhead by requiring specialized vocabularies and rigid grammar structures, enforcing protocols such as the NATO phonetic alphabet and introducing subtle timing errors. Fundamentally, such approaches do not accommodate natural speech variations. Recent LLM-powered tools (e.g. MathGPT Pro, Mathful, StudyFetch) demonstrate improved contextual understanding in highly structured tutoring contexts, but do not offer freeform mathematical authoring \cite{mathgptpro2025, mathful2025, studyfetch2025}. 

In our research, the closest existing tool to \textit{Phoenix} is \textit{EquatIO} \cite{texthelp2025equatio}, which enables robust voice input \cite{o2023trial}. Although existing studies did not rigorously test this feature, as a regular user, P noticed it often created incomplete translations, demonstrating limited contextual understanding (see \autoref{appendix:tests} for test details). Ultimately, while \textit{EquatIO} is useful and feature-rich, it seemingly treats voice as one input modality among many. 

We introduce \textit{Phoenix}, a mathematical input system designed for FMD users to achieve high levels of flexibility and robustness, enabling new levels of abstract problem-solving. By combining LLM-powered contextual understanding with speech recognition, we eliminate the need for specialized vocabularies, enabling an intuitive authoring experience.

\section{Design Procedures and Goals}
\label{sec:design_procedures_and_goals}

\subsection{Design Procedures}
\label{subsec:design_procedures}


We applied the interdependence framework to our mixed-ability co-design process \cite{bennettInterdependenceFrameAssistive2018}. Our team consists of four researchers: one researcher, P, with a FMD and three researchers without an FMD (including one with visual impairment). P utilized English voice dictation software for 5 years, encountering limitations with commercial math dictation tools. Each member of the team has over 5 years of experience with LaTeX.

Our work consisted of bi-weekly online testing sessions from January 2025 to June 2025, during which P initially served primarily as a tester. These sessions focused on overarching goals and identifying areas to streamline Phoenix. Specific activities included testing low- and medium-fidelity prototypes through role-playing, conducting usability testing to refine features based on P’s insights, and drawing on our broader experiences to creatively close gaps. Feedback was gathered through group discussions and recorded think-aloud sessions, ensuring P’s experiences directly shaped usability and design decisions. This feedback was then iteratively incorporated by adjusting prototypes, which we later validated in co-design workshops. In these workshops, P’s role expanded from evaluating designs to actively generating them. Together, we brainstormed features, sketched new interface ideas, and explored alternative interaction flows, using insights from previous testing sessions as a foundation for the next iteration. This process resulted in a functional prototype by June. 

\subsection{Design Goals}
\label{sec:design_goals}

Our fundamental philosophy is to harness conceptual ability through hands-free entry, reducing cognitive load. We aim to create space for deeper mathematical engagement by minimizing fine-motor input and its accompanying obstacles. Our design sessions – grounded in this philosophy – gave rise to four key design goals (DGs). 

\textbf{DG1: \textit{Phoenix} should interpret a wide range of spoken mathematical expressions, symbols, and notation.} Existing tools require users to learn narrow dictation rules, undermining natural speech variations. P expressed how frequently inaccurate transcriptions interrupted workflow, hindering full engagement in high-level mathematics and slowing progress. Thus, we aimed to interpret variability in spoken mathematics, reliably recognizing mathematical dictation in real-world settings.

\textbf{DG2: \textit{Phoenix} should enable flexible equation editing and exporting.}
\textit{Phoenix} should allow various forms of equation editing without navigating complex commands or interfaces. P expressed the need for voice-powered equation editing and exporting work to various platforms like Microsoft Word. Thus, we aimed to enable various forms of equation editing and exporting to change variables or share work, ultimately improving problem-solving. 

\textbf{DG3: \textit{Phoenix} should intuitively convey logical flow of mathematical relationships.} Mathematical problem-solving resembles a web, as complex problems are broken down into smaller, interrelated subcomponents \cite{gilmore_understanding_2023}. Supporting this process, \textit{Phoenix} should help users visually map ideas, making solutions easier to understand and manage. To this end, we aimed to make \textit{Phoenix} both customizable and self-organizing.

\textbf{DG4: \textit{Phoenix} should be flexible to the user’s problem-solving style.} \textit{Phoenix} offers structure for many basic problem-solving features. P expressed how most helpful tools allowed flexibility over rigid usage patterns. Thus, we aimed to maximize flexibility by allowing users to personalize and customize \textit{Phoenix}'s behavior.

\section{System Implementation}
\label{sec:system_implementation}

\begin{figure*}
    \centering
    \includegraphics[height=0.37\textwidth]{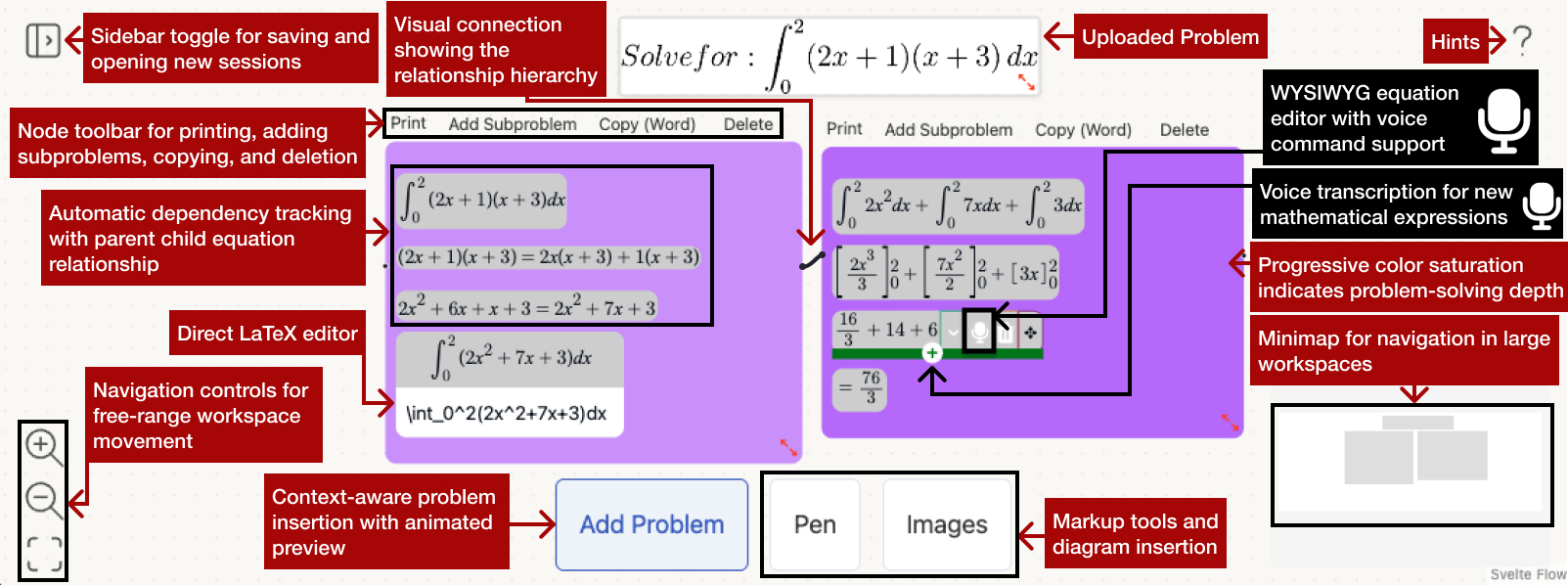}
   \Description{System Overview. A screenshot of the math problem-solving workspace, with boxes and arrows labeling various features. Elements are as follows: A box highlights a question mark icon with the caption "Hints"; Another section points towards the equation editor with the caption "WYSIWYG equation editor with voice command support"; Below this a arrow points towards a plus icon with the caption "Voice transcription for mathematical expressions"; Another box highlights the node background color with the caption "Progressive color saturation indicates problem-solving depth"; Another box points to the mini-map with the caption "Minimap for navigation in large workspaces"; A box highlights the uploaded problem image, with the caption "Uploaded Problem"; A box highlights the pen and images button with the caption "Markup tools and diagram insertion"; A box highlights the visual connectors between nodes with the caption "Visual connection showing the relationship hierarchy"; A label points towards a sidebar toggle icon with the caption "Sidebar toggle for saving and opening new sessions"; A label highlights the node toolbar with the caption "Node toolbar for printing, adding subproblems, copying and deletion"; A box points towards the equations with the caption "Automatic dependency tracking with parent child equation relationship"; A box highlights the equation LaTeX editor option with the caption "Direct LaTeX editor"; A box points towards three icons with the caption "Navigation controls for free-range workspace movement"; A label points towards the add problem button with the caption "Context-aware problem insertion with animated preview".
   }
   \caption{System Overview Diagram}
   \label{fig:system}
\end{figure*}

To address the design goals established during our co-design sessions, we developed a prototype of the \textit{Phoenix} system. In this section, we first provide a detailed explanation of end-user system interaction, followed by a technical description of system architecture. Figure \ref{fig:system} provides a high-level overview of the front-end interface. A detailed evaluation of \textit{Phoenix}’s voice transcription accuracy using benchmark prompts is available in \autoref{appendix:phoenix}.

\subsection{Overview}
\label{subsec:system_overview}
\textbf{Context Engine (addressing DG 1)}
Our most important innovation is the context engine, which transcribes the user's equation while borrowing context from previous transcriptions (including variable names and equation fragments), well-known equations (e.g. the quadratic equation), and domain-specific vocabulary within STEM. Further, this engine accurately transcribes high-level STEM terminology including integrals, partial derivatives, and Greek letters. Ultimately, this engine achieves at least 3 advanced capabilities: (A) Voice-powered equation editing through commands such as “change the plus to a minus”, “substitute 'x' with 7”, or “move the denominator to the numerator”, (B) rendering varied expressions, such as “x over 3,” “x divided by 3,” and “x by 3,” to all yield the same final expression, and (C) isolating math from conversational speech; if a user dictated "I was thinking about the integral of e to the negative x squared", only the math is transcribed.

\textbf{Self-Contained Subproblem Nodes (addressing DG2)} In \textit{Phoenix}, each subproblem forms a self-contained equation space. New subproblems start empty, with just a plus button to transcribe the first equation. An additional collapsible toolbar enables: print the subproblem, add a connected subproblem, delete the node, and copy the contents in Word or LaTeX format, complete with optional step-by-step annotations. Each equation is fully movable and arranged according to user preference (top-to-bottom by default). Equations can be edited by voice, with MathQuill (WYSIWYG) and LaTeX as fallbacks. 


\textbf{Graph-Based Workspace Design (addressing DG3)}
Complex problems contain independent subcomponents, and successful problem solvers use ample space to map ideas in a sensorimotor web \cite{Abrahamson_2016}. Since P expressed the importance of visually mapping equation subproblems, \textit{Phoenix} is graph-based, separating nodes and using color saturation to depict problem hierarchy. For example, a user may dictate, "integral from zero to two of  \((2x+1)(x+3)\)," which the system parses into a root node. Additional nodes are added as users integrate step-by-step, allowing modular, branching thinking. Animated actions reduce visual complexity; hovering over “Add Problem” previews subproblem location.

\textbf{Digital Thinking (addressing DG4)}
Beyond transcription, \textit{Phoenix} supports active, non-linear mathematical thinking. Features like diagram/image uploads allow users to incorporate problems from other platforms. Pen-markup contains various pen colors and thicknesses, enabling creative interaction. Users can also save and load work for future reference. 

\subsection{Architecture}
\label{subsec:system_architecture}
\begin{figure*}
    \centering
    \includegraphics[height=0.3\textwidth]{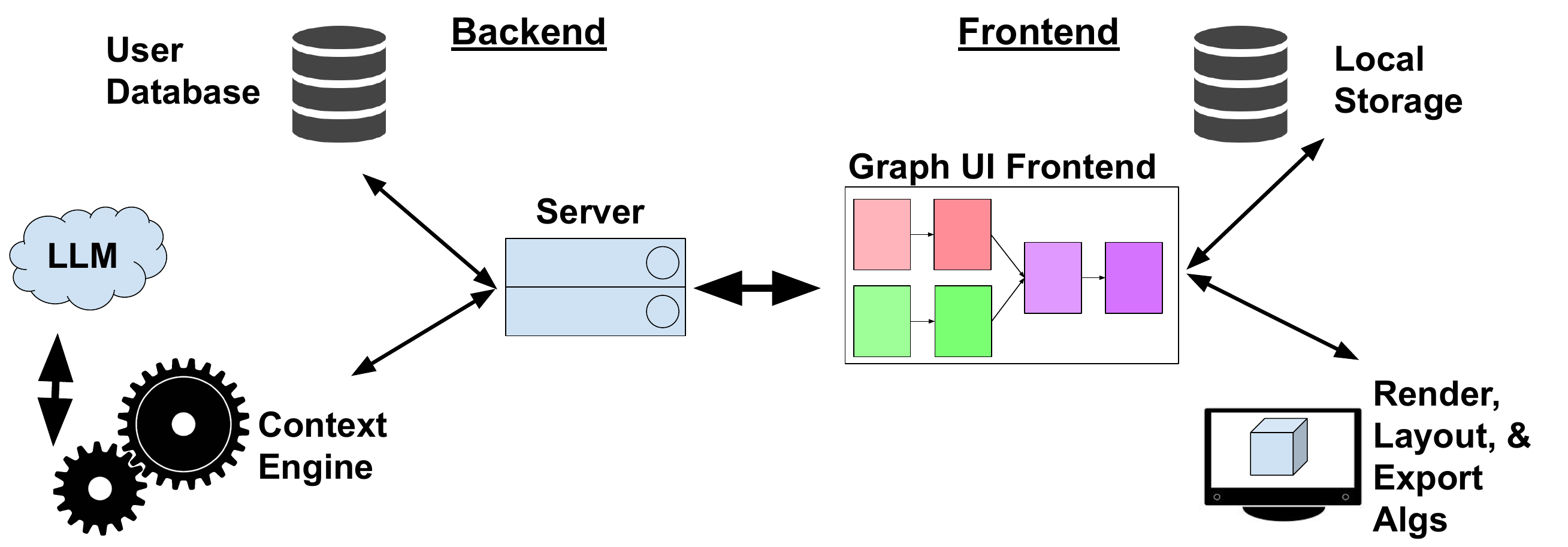}
   \Description{Architecture Overview. On the left side is the backend, and on the right side is the front end. On the backend, the server connects to the user database and the context engine. The context engine is also connected to the LLM. On the front end, the graph UI frontend connects to both local storage and rendering, layout, and export algorithms. The server on the backend passes data to and from the graph UI frontend.}
   \caption{Architecture overview, showing the flow of data from the graph-based frontend to the backend, context engine, and beyond}
   \label{fig:architecture}
\end{figure*}
We built \textit{Phoenix} with a Svelte frontend, a context engine that builds LLM queries, and a Node.js backend. 

\textbf{Client Side Data Structures} We use persistent storage to store substantial data on the client side, such as images, markup, documents, and user preferences. We designed a top-down data structure to store subproblem nodes, including a custom JSON-based data structure tracking equation data and positioning, background color, and size. We used a robust vector-based approach for markup, saving a list of individual paths (rather than pixels), rendering only those visible onscreen. Finally, we store user preferences on the client side in a persistent store. 

\textbf{Export Algorithms} To help users share and export their work, we developed a "print" function, and integrations to copy work into LaTeX and Microsoft Word. The print feature loads a new popup with all equations for a given subproblem in a tabular format, renders them using MathJax, and automatically triggers a system print dialogue upon successful render. We built a one-click button to copy annotated steps directly in LaTeX, formatting equations into a specially designed template, leveraging the system clipboard. 

The most challenging export algorithm was integrating with Microsoft Word. Word uses a restricted subset of MathML, normally making it impossible to copy more than one equation to the clipboard (this will appear as text, rather than an equation object). To overcome this barrier, we leveraged the table feature to copy multiple equations at once, using ‘mtext’ elements to annotate steps. 

\textbf{Server Side User Management} Our backend implements authentication using Google OAuth and Json Web Tokens (JWTs), managing server-side user info for the context engine, and rate limiting to ensure fair distribution of limited API queries. To further ease backend load, we use the Web Speech API to transcribe speech into text, before sending the text to the context engine to generate equations. Unlike prior work, we do not need a custom speech-to-text model because our context engine flexibly infers important details and correct mistakes. 

\textbf{Context Engine} The context engine is applied whenever the user edits an equation or generates a new one via voice (i.e. transcription or transformation). Context is managed as soon as the user interacts with the page. In addition to the overt dependencies \textit{between} subproblems, \textit{Phoenix} also maintains an implicit dependency order among the equations \textit{within} each subproblem. When users transcribes a new equation, \textit{Phoenix} automatically tracks its “parent” equation on creation. All this information is sent to the context engine. 

Using these dependencies, the engine runs a graph traversal to order equations from most to least relevant. This graph traversal calculates a distance-weighted “fuzzy usefulness” to weight the context of each step. Unlike approaches like RAG, our graph traversal retains full detail, is more interpretable, and allows greater control over the subject domain.

After ordering the dependencies, we attach them to our LLM prompt, instructing the model to consider most relevant context. To generate transcriptions, we use few-shot prompting by providing example inputs and outputs to help the model understand how to handle complex commands and queries involving domain-specific knowledge (e.g. physics terminology). The result is then retrieved and sanitized by a regular expression before being sent back to the client. 

\section{Future Directions}
\label{sec:futuredirs}
Our next step is focused on helping \textit{Phoenix} more fully realize our design goals. Future work will target universal voice control,  improving occasionally inaccurate complex STEM equation transcription (e.g. quantum mechanics), respecting user preference for default lowercase letters, and enabling scroll-based navigation over click-and-drag interaction. We are excited to gather feedback in our upcoming user study.

We have also released a public beta of \textit{Phoenix}, free for users to try, at \url{https://assistivity.org}.

\begin{acks}
  This project did not receive any external funding, and was entirely done out of the passions of our coauthors. We wanted to especially acknowledge Kenneth Ge's contributions with self-funding and leading the project, Ryan Paul's contributions with designing interfaces and leading workshops, and Priscilla Zhang's contributions with providing creative ideas that helped ground our work. 
\end{acks}

\bibliographystyle{ACM-Reference-Format}
\bibliography{references/extra}

\appendix
\section{Detailed Test Conditions and Results}
\label{appendix:tests}

\subsection{Test Environment}
\begin{itemize}
    \item \textbf{Device:} Windows laptop
    \item \textbf{Browser:} Google Chrome
    \item \textbf{EquatIO Version:} Free version (\textit{Accessed June 12, 2025)}
    \item \textbf{Access Method:} Via redirect link \url{https://equatio.texthelp.com} (sourced through ChatGPT)
    \item \textbf{Installation Process:}
    \begin{itemize}
        \item Installed Chrome extension and desktop application.
        \item Attempted sign-in via desktop app, which redirected to a browser login page that repeatedly failed to load.
        \item Signing in with Google also led to non-loading page.
        \item Ultimately accessed EquatIO through the redirect link above; this link was not easily discoverable through standard web search.
    \end{itemize}
\end{itemize}

\subsection{Testing Process - EquatIO}

Equations were dictated using EquatIO's speech input feature. Below are the prompts used and their corresponding outputs. 

\subsubsection*{Prompt 1: ''The integral from zero to infinity of x squared, dx''}
\begin{itemize}
  \item \textbf{Expected Output:}
  \[
  \int_0^{\infty} x^2 \, dx
  \]
  \item \textbf{Actual Output:}
    \begin{figure}[H]
    \includegraphics[width=0.2\textwidth]{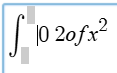}
   \Description{Prompt 1 \textit{EquatIO} Response Overview. Screenshot of an incorrectly formatted integral generated by \textit{EquatIO} after voice dictating, 'The integral from zero to infinity of x squared, dx.' The integral symbol appears with misplaced limits: ‘0’ and ‘2of’ are shown inline instead of as proper lower and upper bounds. The phrase ‘infinity’ was not recognized, and the limits were not formatted correctly.}
   \end{figure}
  \item \textbf{Notes:} Integral notation and limits were not recognized; the output was a partial, incorrect transcription.
\end{itemize}

\subsubsection*{Prompt 2: ''The integral from zero to pi over two, cosine of x, dx''}
\begin{itemize}
  \item \textbf{Expected Output:}
  \[
  \int_0^{\frac{\pi}{2}} \cos(x) \, dx
  \]
  \item \textbf{Actual Output:} 
    \begin{figure}[H]
    \includegraphics[width=0.3\textwidth]{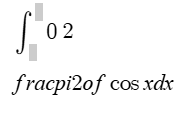}
   \Description{Prompt 2 \textit{EquatIO} Response Overview. Screenshot of an incorrectly formatted integral generated by \textit{EquatIO} after voice dictating, 'The integral from zero to pi over two, cosine of x, dx.' The integral incorrectly shows the limits ‘0’ and ‘2’ vertically beside the integral sign. The integrand contains the malformed expression ‘fracpi2of cos x dx’, indicating that the phrase ‘pi over two’ was misinterpreted as ‘frac pi 2 of’ and not properly typeset.}
    \end{figure}
  \item \textbf{Notes:} Similar issues with parsing. Fractional upper limit and function notation were partially recognized but inaccurately rendered.

\end{itemize}

\subsubsection*{Prompt 3: ''Index of refraction one sine of theta one equals index of refraction two sine of theta two''}
\begin{itemize}
  \item \textbf{Expected Output:}
  \[
  n_1 \sin(\theta_1) = n_2 \sin(\theta_2)
  \]
  \item \textbf{Actual Output:}
  \begin{figure}[H]
    \includegraphics[width=0.3\textwidth]{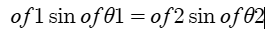}
   \Description{Prompt 3 \textit{EquatIO} Response Overview. Screenshot of an incorrectly formatted Snell's Law equation generated by \textit{EquatIO} after voice dictating 'Index of refraction one sine of theta one equals index of refraction two sine of theta two'. The output incorrectly displays ‘of1 sin of θ1 = of2 sin of θ2’, where the intended variable ‘n’ for index of refraction was misrecognized. The formatting is otherwise close, but key variables are incorrect.}
    \end{figure}
  \item \textbf{Notes:} \textit{EquatIO} did not associate “index of refraction” with the variable $n$.
\end{itemize}

\subsection{Testing Process - \textit{Phoenix}}
\label{appendix:phoenix}

The same prompts were dictated using \textit{Phoenix}’s speech input feature in the identical test environment. The outputs below were generated using the latest prototype version of \textit{Phoenix} as of June 2025.

\subsubsection*{Prompt 1: ''The integral from zero to infinity of x squared, dx''}
\begin{itemize}
  \item \textbf{Expected Output:}
  \[
  \int_0^{\infty} x^2 \, dx
  \]
  \item \textbf{Actual Output:}
    \begin{figure}[H]
    \includegraphics[width=0.25\textwidth]{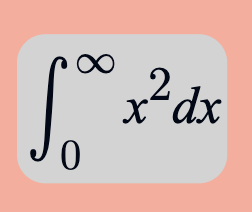}
    \Description{Prompt 1 \textit{Phoenix} Response Overview. Screenshot of a correctly formatted integral generated by \textit{Phoenix}. The integral symbol includes properly formatted lower and upper limits from 0 to infinity, and the integrand is correctly shown as x squared dx.}
   \end{figure}
  \item \textbf{Notes:} \textit{Phoenix} successfully recognized and formatted the integral expression, including both bounds and the integrand.
\end{itemize}

\subsubsection*{Prompt 2: ''The integral from zero to pi over two, cosine of x, dx''}
\begin{itemize}
  \item \textbf{Expected Output:}
  \[
  \int_0^{\frac{\pi}{2}} \cos(x) \, dx
  \]
  \item \textbf{Actual Output:}
    \begin{figure}[H]
    \includegraphics[width=0.25\textwidth]{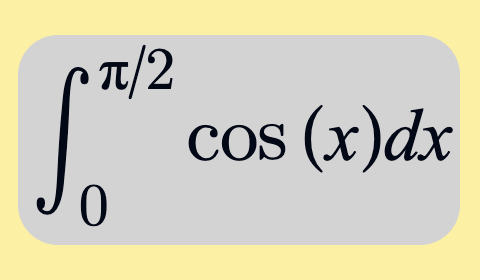}
    \Description{Prompt 2 \textit{Phoenix} Response Overview. Screenshot of a correctly formatted integral generated by \textit{Phoenix}. The upper limit appears as pi over two, rendered as a proper LaTeX fraction, and the integrand is written as cosine of x dx.}
   \end{figure}
  \item \textbf{Notes:} The system accurately parsed both the limits and the function. “Pi over two” was correctly rendered as a LaTeX fraction.
\end{itemize}

\subsubsection*{Prompt 3: ''Index of refraction one sine of theta one equals index of refraction two sine of theta two''}
\begin{itemize}
  \item \textbf{Expected Output:}
  \[
  n_1 \sin(\theta_1) = n_2 \sin(\theta_2)
  \]
  \item \textbf{Actual Output:}
    \begin{figure}[H]
    \includegraphics[width=0.25\textwidth]{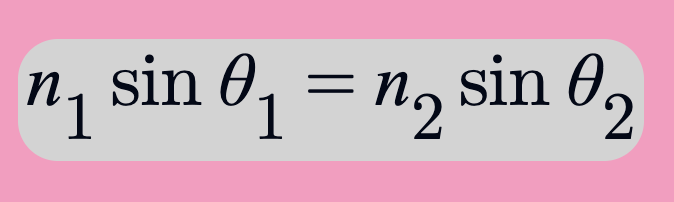}
    \Description{Prompt 3 \textit{Phoenix} Response Overview. Screenshot of a correctly formatted version of Snell’s Law generated by \textit{Phoenix}. Variables n1 and n2 were used correctly to represent indices of refraction, and sine and theta were accurately rendered.}
   \end{figure}
  \item \textbf{Notes:} Unlike EquatIO, \textit{Phoenix} correctly mapped “index of refraction” to $n$ and handled symbolic input well.
\end{itemize}

\end{document}